(IJCSIS) International Journal of Computer Science and Information Security,
Vol. 4, No. 1, 2009# Computational Complexities and Breaches in Authentication Frameworks

## of Broadband Wireless Access (BWA)

Raheel Maqsood Hashmi [1], Arooj Mubashara Siddiqui [2], Memoona Jabeen [3]
Khurram S. Alimgeer, Shahid A. Khan
Department of Electrical Engineering
COMSATS Institute of Information Technology Islamabad, Pakistan
[1] rahilmh@gmail.com, [2] aroojmubashara@gmail.com, [3] memoona.jabeen@gmail.com*Abstract*— Secure access of communication networks has become an increasingly important area of consideration for the communication service providers of present day. Broadband Wireless Access (BWA) networks are proving to be an efficient and cost-effective solution for the provisioning of high rate wireless traffic links in static and mobile domains. The secure access of these networks is necessary to ensure their superior operation and revenue efficacy. Although authentication process is a key to secure access in BWA networks, the breaches present in them limit the network's performance. In this paper, the vulnerabilities in the authentication frameworks of BWA networks have been unveiled. Moreover, this paper also describes the limitations of these protocols and of the solutions proposed to them due to the involved computational complexities and overheads. The possible attacks on privacy and performance of BWA networks have been discussed and explained in detail.

*Keywords- Comutational Complexity; Authentication; Security; Privacy; Key Management.*## I. INTRODUCTION

Broadband Wireless Access (BWA) is rapidly emerging as the standard for future communication networks. The ease of deployment combined with low operational and maintenance costs makes BWA the preferred choice for modern communication service providers. The BWA or WiMAX (World-wide Interoperability for Microwave Access) networks work on the protocols defined in the IEEE 802.16 standard [1]. IEEE 802.16 has two revisions: 802.16d termed as fixed WiMAX and 802.16e termed as mobile WiMAX [2]. The deployments of WiMAX networks are growing rapidly to achieve seamless mobility followed by worldwide broadband communications.

Authentication of users and of equipment in the BWA network is done as a part of the admission control process. The authentication phase is also carried out while execution of handoffs in mobile BWA networks. The authentication and service authorization process is carried out at the privacy sub-layer, embedded in the WiMAX protocol stack [1], [3]. A complete protocol ensuring secure distribution and management of keying data between network entities is incorporated in this layer, known as Privacy and Key Management protocol (PKM) [1]. Launch of 802.16d in 2004 and 802.16e in 2005 suggests that the standard is in the initial phase of implementation and several dormant issues and short comings will be highlighted with progress in deployment and service provisioning.

Network security and legitimate service access is a concealed performance indicator in providing Quality of Service (QoS) to users. In this paper, the pitfalls in the current authentication frameworks have been unveiled, the reasons for the breaches have been identified and the causes, been analyzed to highlight the limitations of the existing protocols.

Rest of the paper is organized as follows. In Section II, we introduce the existing authentication frameworks. Section III describes the attacks on authentication. Section IV highlights the computational complexities and overheads involved in the existing protocols and Section V concludes our discussion.

## II. AUTHENTICATION FRAMEWORKS

### A. Privacy & Key Management Protocol version 1:

The PKM v1 protocol complies with the 802.16d-2004 standard and is operating in the Fixed WiMAX networks. This protocol is a 3-step protocol involving 1-way authentication. The figure 1 shows the PKM v1 authentication model and messages involved.

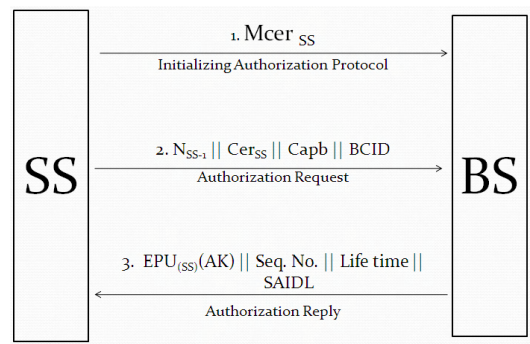

Figure 1. Privacy and Key Management Protocol version 1 [5]

The detailed operation of PKM v1 can be found in [1], [4] and [5]. PKM v1 is based on X.509 certificate based Public

ISSN 1947 5500

Key Infrastructure (PKI). Figure 1 shows the information flow between Subscriber Station (SS) and Base Station (BS). The individual components of the message have been addressed in [1] and [5]. In step 2, a nonce ($N_{SS}$) is shown which is a 64-bit number generated randomly to be used as a message linking token [4]. Basic Connection Identity Code (BCID) is used to identify a particular node in the network and is assigned to the node during the admission control process.

*B.  Privacy &Key Management Protocol version 2:*

PKM v2 protocol was defined in 802.16e-2005 and is implemented in Mobile WiMAX networks. This protocol is not essentially a variant of PKM v1. However, PKM v1 and v2 share a common service authorization structure. PKM v2 is a 4-step, 3-way authentication protocol. The operational mechanism of PKM v2 is illustrated in [2] and [6]. Figure 2 depicts the PKM v2 authentication framework.

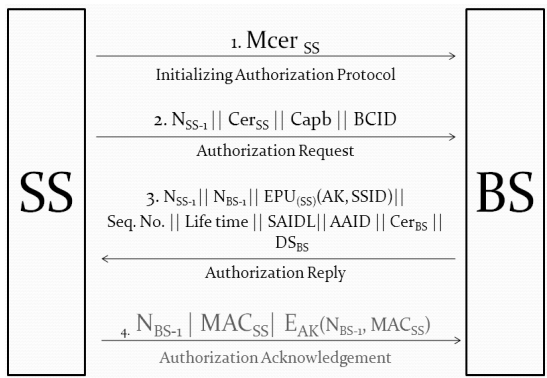

Figure 2. Privacy and Key Management Protocol version 2 [6]

The major enhancements in PKM v2 are the inclusion of digital certificates (DS) and authorization acknowledgement step. Moreover, except step 1, a nonce ($N_{SS}$ or $N_{BS}$) has been incorporated with each message to link successive steps of the protocol.

*C.  The Time Stamp Authorization (TSA) Model:*

This model has been proposed by Sen Xu et al. in [7] and introduces timestamps to ensure message authenticity. This proposal is a variant of PKM v1 which has timestamps placed on all messages to certify the freshness of the message. Each node in the network (BS or SS) maintains a timestamp table which contains the timestamps of all the messages received; therefore, preventing the message replays. Furthermore, a key management strategy for inter-BS key management and exchange has also been proposed along with this protocol in [7]. This model specifically focuses and enhances PKM v1 authorization model and is aimed for fixed WiMAX networks.

*D.  Hybrid Authorization (HA) Model:*

Ayesha Altaf et al., in [6], propose a model which employs a hybrid approach involving nonce and timestamps to prevent the attacks on privacy and key management protocols. This proposal claims to cater the effect of interleaving attacks discussed in [7] and [8] which may occur on PKM v2. The approach is presented for mobile WiMAX networks and enhances the PKM v2 authentication framework.

*E.  Improved Secure Network Authentication Protocol:*

This model has been proposed in [4] and aims to restructure the authentication framework by introducing a single protocol for both fixed and mobile networks. It has been introduced to cover some of the major threats highlighted in [6], [7] and [9]. Improved Secure Network Authentication Protocol (ISNAP) has been designed and optimized by utilizing the existing system resources involving minimum overhead. The proposed model of ISNAP is shown in figure 3.

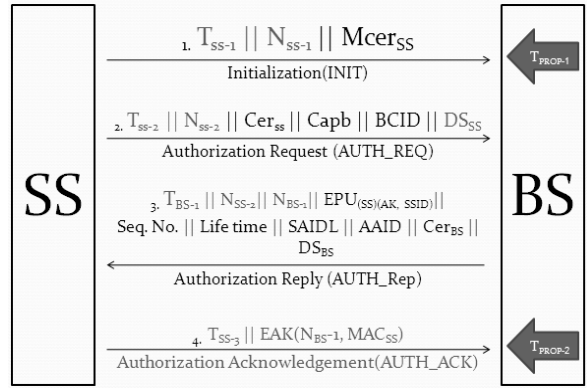

Figure 3. Improved Secure Network Authentication Protocol [4]

The detailed structure and working of ISNAP has been discussed in [4].

### III.  ATTACKS ON AUTHENTICATION

Attacks on authentication can be described as the ways by which a network can be intruded and the privacy of the users be compromised. The secure access of network services is becoming an increasingly important issue in the present communication infrastructures. Any attempts of an interloper to get registered with the network illegitimately or to create chaos in it, is possible; if the user authentication and authorization stage is compromised. Therefore, the ways to breach the authentication frameworks are termed as attacks on privacy and key management protocols and their variants.

*A.  Water-Torture Attack:*

The Water-Torture attack is aimed to perturb the network's operation by causing flooding. There are some messages which are used to initiate cyclic processes when received on any node. Transmission of such a message can be seen in figure 1 and figure 2 as $Mcer_{SS}$. $Mcer_{SS}$ is the manufacturer's X.509 certificate which is used by the SS to show its presence in the network and to initiate the authentication protocol [5]. In the admission control process, the reception of this message at BS initiates the cyclic authentication procedure. In the event of a Water-Torture attack, these triggering messages are captured and are transmitted in a loop to cause trigger flooding; thus, creating artificial congestion at the BS. This attack can be extended to create a management chaos and blocking in the network's domain, especially, where remote power feeding is employed. PKM v1 & v2, TSA Model and the HA model are influenced by this attack as there is no method to detect if Mcer$_{SS}$ has been transmitted by an authentic SS or has been replayed by an intruder.

*B. Denial of Service Attack:*

While dealing with radio frequency networks we come across certain entities which maintain a cause and effect relationship between them. The Denial of Service (DoS), in this case, is one of the results of the Water-Torture attack. Due to the engaged resources and falsified congestion, the requests for authentication, key exchange and securing admission to the network are not entertained. This causes a severe degradation of QoS, therefore, resulting in heavy revenue losses. The protocols subjected to Water-Torture attack are, naturally, also subjected to DoS attack.

*C. Message Replay Attack:*

This attack, under its footprint, covers a large number of intrusion methods which are based on the employment of the described approach. This attack involves the capturing and re-use of the messages in the authentication cycles. The re-use can be based on a specific message or on a set of messages exchanged during a complete session. PKM v1 is not supported with any mechanism to counter this attack. However, PKM v2 partially counters this attack by employing nonce in message 2 and 3 as shown in figure 2. Nonce being a 64-bit random number has $(2^{64})^{-1}$ probability of repetition and is very difficult to be predicted. It does prove useful to link subsequent messages and helps to resolve the replay issues to a partial extent. Hence, PKM v1 is a victim of replay attacks while PKM v2, partially not completely, is secure. The TSA model proposes timestamps instead of nonce while the HA model demands the use of nonce in conjunction with timestamps, but both models present significant overhead, as discussed in next section.

*D. Identity Theft:*

The SS equipment in the network is provisioned with the services on the basis of the physical (MAC address) registered in the network. In case of fixed BWA networks, the MAC identities are registered permanently for each SS, however, in mobile BWA networks, the MAC ID is registered each time a node joins the network or performs handoffs. Hence, PKM v1 is not exposed to this attack, but, as in figure 2, PKM v2 and HA model are vulnerable to this attack as message 4 contains MAC identity in both encrypted and unencrypted form. There are several devices available at the present day which can be reprogrammed with variable MAC addresses [6], [9].

*E. Impersonation:*

Impersonation refers to the type of attack in which one node masquerades another node. There are several ways in which impersonation can be achieved like by message replay or while employing one-way authentication [12]. The PKM v1 model and Time-Stamp authentication model are vulnerable to this type of infringement attempt. The reason for this is one-way authentication i.e., BS authenticates SS but vice versa does not occur. Moreover, this attack is aimed to compromise the security of the users and poses severe threats in case of employment of BWA infrastructure in security and defense installations for any realm.

*F. Interleaving Attack:*

Interleaving attack is a sub-class of Man-in-the-Middle attacks and is specifically aimed for PKM v2. In this attack, an adversary interleaves a communication session by maintaining connections with the BS and SS, pertaining as SS to BS and vice versa. All the information on route passes through the adversary node and thus an information leakage point is built [8]. The backbone of interleaving attack is the re-transmission of a set of messages from the same session. The HA model proposes an approach to cater the interleaving attack by introducing transmission and storage overheads in the network [6].

*G. Suppress Replay Attack:*

This method of gaining forged access to the network services takes advantage of the fact that perfect synchronization must be maintained to protect the authentication session from intrusion. Due to the loss of synchronization in the clocks of the entities, an intruder can gain control on the authentication framework by capturing the messages and transmitting them with added delays, thus causing forward message replay [6]. This class of attack is difficult to counter and is vulnerable for the Timestamp Authentication model. The Hybrid Authentication model can also be manipulated by this attack.

IV. COMPUTATIONAL COMPLEXITIES AND OVERHEADS

The Timestamp Authentication model, Hybrid Authentication Model and ISNAP have been put forth to remove the threats posed to the standardized protocols PKM v1 and PKM v2. The first two models focus their predecessors i.e. PKM v1 and PKM v2, respectively, for removal of threats; however, ISNAP focuses on a single solution for fixed and mobile BWA networks, solving the existing problems. The proposed models, along with the enhancements, offer computational complexities and storage overheads as discussed in this section.

*A. Timestamp Validation:*

The TSA model, HA model and ISNAP model have been put forth to remove the threats posed to the standardized protocols PKM v1 and PKM v2. The first two models focus their predecessors i.e. PKM v1 and PKM v2, respectively, for removal of threats; however, ISNAP model focuses on a single solution for fixed and mobile BWA networks, along with eradication of the posed threats.

Timestamps are freshness guards for messages. These timestamps, used to eliminate replays, are recorded in timestamp tables. These tables contain the timestamps for all the messages previously received and are used to compare the timestamp of any newly received message with the recorded ones. The presence of the newly received timestamp in the tables leads towards the detection of a replay, or otherwise, a validation of message.

The timestamp tables consume memory for the storage of prior timestamps from messages and also consume a large number of computational cycles to compare the arriving timestamp with the recorded ones. Let $\partial$ be the number of bytes in the memory for storage of one timestamp and $\rho$ be the number of days for which the records are maintained in the table. Then we have in (1) as:

$$\chi := \Psi \times \partial \times \rho \quad (1)$$

where $\chi$ is storage overhead caused by the timestamp tables expressed in bytes/node and $\Psi$ is the minimum number of messages exchanged between two communicating nodes per day. Generally, to counter the replays, timestamp records are maintained in the tables for an optimum amount of time. Thus, assuming $\partial$ and $\rho$ to be 4 bytes (as in a UNIX based environment) and 15 days, if $\Psi$ be a minimum of 100 messages validated per day, the value of $\chi$ approaches to 6 Kilobytes to be maintained for each node. Hence for a BS serving 64 SSs, this can lead to 0.3 Megabytes of minimum static memory reserved by timestamp tables at each BS.

A very general implementation of timestamp comparison on UNIX based operating system suggests that a minimum of 2 floating point instructions are used for comparison of one timestamp. Therefore, the machine cycles for comparison of the timestamps can be calculated by (2) as:

$$\alpha = 2^\chi. \quad (2)$$

where $\alpha$ is the number of computational cycles used in the timestamp validation process. Thus, we can have the number of floating point instructions per second (FLOPS) in (3) as:

$$\text{FLOPS} = \alpha \times (\sigma)^{-1} \quad (3)$$

where $\sigma$ is the number of machine cycles per second for any particular system (SS or BS). The above analysis suggests that the number of FLOPS used in the timestamp validation process will be significantly large depending upon the amount of records maintained and the constriction time required to counter replay attacks. The final expression for the number of FLOPS becomes:

$$\text{FLOPS} = 2^{(\rho.\partial)} \times (\sigma)^{-1} \quad (4)$$

The above analysis suggests that the storage overhead is quite significant in terms of reserving the memory resources for any system and can be optimized by enhancement of the timestamp table comparison method. The proposed models to rectify the threats discussed in section III, the TSA model and the HA model, are subject to severe limitations discussed above. ISNAP model, however, offers the replacement of the timestamp table method by offering a validation procedure based on mathematical evaluations [4]. In this case, a timestamp is subjected to a mathematical condition; if the condition is fulfilled, the message is validated, else the message is contradicted. Therefore, ISNAP's validation procedure reduces the storage overheads to very considerable limits by removing the need for maintaining record tables.

*B. Transmission Overheads:*

In order to minimize the posed threats, different sentinels are to be introduced in the authentication frameworks. This increases the transmission overhead for the verification procedures; therefore, establishing a cost and benefit relationship between the security and increased transmission.

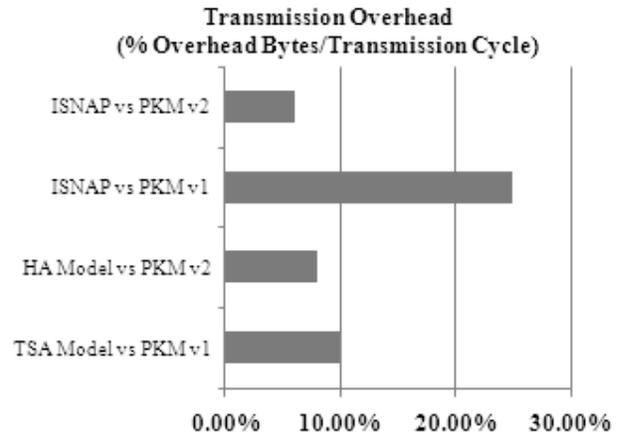

Figure 4. Transmission Overheads

Figure 4 shows the comparative transmission overheads of the proposed models with their standardized counterparts. ISNAP model poses lesser transmission than Hybrid Authentication model by removing some redundant components like unencrypted MAC ID in the last message. As for the TSA model, as it is a variant of PKM v1, the fixed network protocol, ISNAP requires substantially more transmission. However, the cost and benefit relation is justified due to the removal of several major attacks in ISNAP and reduction in storage resources.

Combined with overheads involving storage resources and transmission overheads, the maximum operating cost is offered by the proposed HA model and afterwards TSA Model and ISNAP model have comparable outlays. However, based on the performance of these proposed solutions, ISNAP provides optimum protection against intrusion and unauthorized use of network resources.

*C. Resynchronization Scheme:*

The involvement of clock based parameters in authentication procedures suggests that the clocks of the involved systems must be well synchronized to allow for a successful handshake between the intervening nodes. The synchronization scheme in WiMAX networks has remained an issue as discussed in [10] and there remains probability of errors in the clocks [11]; allowing intrusion activity. Although the ISNAP model suggests a solution to this issue, the analysis and implementation of the synchronization schemes has yet to be performed to reach a satisfactory conclusion.

*D. Initiation of Authentication Procedure:*

In PKM v1, PKM v2, Timestamp Authentication Model and Hybrid Authentication model, the trigger message for initiating the handshake procedure cannot be protected against the class of replay attacks. ISNAP model proposes a solution to this vulnerability but demands the clocks to be synchronized.

## V. CONCLUSION

The authentication protocols standardized in the WiMAX or BWA networks are faced with a number of vulnerabilities which are critical to smooth operation of the network and demand shear attention. The proposed solutions to the posed threats have been, up to some extent, successful in sorting out the issues but not feasible enough in terms of the proposed complexities and overheads. However, ISNAP model has been optimum in terms of solving the security issues along with offering optimized use of resources. Nevertheless, optimization of operations for validation procedures and finest use of system resources to furnish secure network access is required and demands more research in this area.

### AUTHORS PROFILE

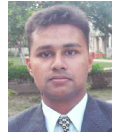
**Raheel M. Hashmi** is a graduate student enrolled in MS Engineering at Politecnico di Milano, Italy. He did his degree in Electrical Engineering from COMSATS Institute of Information Technology (CIIT), Islamabad in 2009 and received Gold Medallion Award. He has research contributions in the area of Wireless Networking and Security.

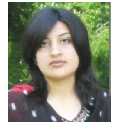
**Arooj M. Siddiqui** has done Electrical (Telecom.) Engineering from Dept. of Electrical Engineering, CIIT, Islamabad in 2009. She is a Graduate Student and Researcher and has contributed towards the area of Authentication in BWA Networks.

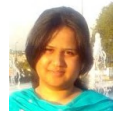
**Memoona Jabeen** completed Electrical Engineering degree from CIIT, Islamabad, Pakistan in 2009. She has International Research Publications in the area of Secure Wireless Access and Cryptographic Methods.

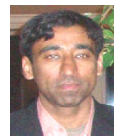
**Khurram S. Alimgeer** did his Bachelors degree in IT in 2002 and completed his MS in Telecommunications with distinction in 2006. He has been with Dept. of Electrical Engineering, CIIT since 2003 and been supervising extensive research work. Currently, he is Assistant Professor at CIIT and is also working as doctoral researcher. His areas of research include Wireless Communications, Image Processing & Antenna Design.

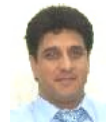
**Professor Dr Shahid A. Khan** did his Bachelors in Electrical Engineering in 1988 from UET Taxila. He did MS Electrical and Electronics Engineering and Ph.D. in Communications from University of Portsmouth, UK. Since then, he has been involved in significant R&D work with research giants like European Antennas Ltd. UK, Newt International Ltd. UK and WAPDA. He joined CIIT in 2003and is, at present, serving as Dean, Faculty of Engineering CIIT. He has significant research contributions to the field of Wireless Networks.